\documentclass[12pt]{article}

\catcode`\@=11

\def\section{\@startsection{section}{1}{\z@}{3.5ex plus 1ex minus
 .2ex}{2.3ex plus .2ex}{\bf}}

\def\thesubsection{\arabic{section}.\arabic{subsection}}
\renewcommand{\subsection}[1]{\addtocounter{subsection}{1}
\vspace{2.5mm}\par\noindent {\it \thesubsection . #1}\par
 \vspace{0.5mm} }
\catcode`\@=12
\newfont{\mbm}{msbm10 scaled\magstep1}

\def\be{\begin{equation}}
\def\ee{\end{equation}}
\def\ba{\begin{eqnarray}}
\def\ea{\end{eqnarray}}
\begin{document}
\def\pct#1{\input epsf \centerline{ \epsfbox{#1.eps}}}

\begin{titlepage}
\hbox{\hskip 12cm ROM2F-87/25  \hfil}
\vskip 1.4cm
\begin{center}  {\Large  \bf   
Open strings and their symmetry groups}

\vspace{1.8cm}
 
{\large \large  Augusto Sagnotti}
\vspace{0.8cm}

{\sl Dipartimento di Fisica\\  Universit{\`a} di Roma 
\ ``Tor Vergata'' \\ INFN, Sezione di Roma \\
Via Orazio Raimondo \\ 00173 \ Roma \ \ ITALY}
\end{center}
\vskip 4.5cm

\begin{center}
{\it Talk presented at Cargese '87, ``Non-Perturbative Quantum Field Theory'', 
July 1987. 
\vskip 12pt \noindent
Published in ``Nonperturbative Quantum Field Theory, eds. 
G. 't Hooft, A. Jaffe, G. Mack, P.K. Mitter and R. Stora (Plenum Publishing
Corporation, 1988), pp. 521-528.} 
\end{center}

\vskip 2.5cm
\begin{center}
{( October , \ 1987 )}
\end{center}
\vfill
\end{titlepage}
\vglue 7cm
{\pagestyle{empty}
{\it The author would like to thank the SPIRES group, and in particular the
Scientific Database Manager Travis Brooks and the previous
Scientific Database Manager Heath O'Connell, for the kind help in linking
this manuscript to the related literature.}
\vfill\eject}
\makeatletter
\@addtoreset{equation}{section}
\makeatother
\addtolength{\baselineskip}{0.2\baselineskip} 

The last three years have seen a large amount of progress in String Theory
\cite{prog}, and the subject itself has undergone a change of scope. This is
well reflected in the content of the talks that have preceded this one.

Much of the recent progress in String Theory can be traced to a precise
strategy: a careful study of the few models known since the beginnings of
the subject, and the abstraction from them of basic properties that one would
like to demand from other models. This could be termed a set of 
``model-building rules''.  The approach corresponds to the fact, often a
source of embarrassment to specialists, that String Theory, born as a set of
rules rather than as a set of principles, has long resisted attempts to 
reduce it to a logically satisfying structure.

The first property is {\bf two-dimensional conformal invariance}. This is
the motive behind the machinery that has become known as {\bf Two-Dimensional
Conformal Field Theory} \cite{cft1,cft2}. Conformal invariance 
entered the subject
long ago, as the cure to the unitarity problems of the bosonic model, by 
necessity defined in terms of oscillators of Minkowski, rather than Euclidean,
signature. Conformal invariance is responsible for a transverse spectrum of
states, all of positive norm, and embodies the projective invariance
(duality) of the originally known amplitudes. Conformal invariance is 
reflected, in the context of the {\bf closed} bosonic string, in the 
invariance under two mutually commuting Virasoro algebras. 

Conformal Field Theory is an algebraic, and thus non-perturbative, framework
for the description of all models that share this property of invariance
under two distinct Virasoro algebras. The algebraic description proceeds
via ``primary fields'' (tensors of the conformal group), and entails the 
encoding of the two-dimensional dynamics in the coefficient 
functions of the operator
product algebra. The operator product coefficients are somewhat reminiscent, in
their role, of the structure constants in ordinary Lie algebras. Like 
structure constants, they are subject to quadratic constraints. These embody
the duality properties of amplitudes. In principle, these data are sufficient
to reconstruct the correlators of the fields, and thus the scattering 
amplitudes of String Theory. The {\bf closed bosonic string} in 26 dimensions
is built out of a particular two-dimensional conformal field theory, one
consisting of 26 copies of a free massless bosonic field. This model has
a distinctive feature: the value of the central charge is 26 for each of the
two independent Virasoro algebras. In modern terms, this compensates the 
contribution (equal to -26) of the conformal field theory of the 
reparametrization ghosts, to give a resulting model which is exactly
{\bf conformally invariant}, {\it i.e.}~characterized  by a total
central charge equal to zero \cite{pol}. Alternatively, this property is
linked to the absence of local anomalies in the two-dimensional field theory.
Generalizations of the bosonic model are to retain this property, if they
are to describe a space time of Minkowski signature, and a suitable 
generalization is needed to deal with (possible) super-partners of the
string coordinates, in the form of {\bf super-conformal invariance}.

The second property is {\bf modular invariance}. It is essentially the
statement that the theory be a theory of closed strings. This property
was identified long ago \cite{mod}, again in the context of the bosonic
model. In modern language, following Polyakov \cite{pol}, one would correct
the tree amplitudes of String Theory by the addition of terms where the 
Conformal Field Theory lives on surfaces of increasing genera. For 
simplicity, let us restrict our attention to the vacuum amplitude, since
this is supposed to capture the essence of the matter anyway \cite{cft2},
and let us consider the first correction to the ``tree'' amplitude. In this
case the parameter space has the topology of the torus. The complex structure
of the torus, the one datum that a truly conformally invariant model feels,
can be described via a two-dimensional lattice, of sides $1$ and $\Omega$,
where one can take $IM(\Omega)$ to be positive (fig. 1). $\Omega$ is the
``period matrix'' of the torus. The crucial point is that doubly periodic
(elliptic) functions, such as the bosonic string coordinates, have to way
to distinguish between the two sides of the cell, and in general between
choices of fundamental cell related by the familiar redefinition
\begin{equation}
\Omega \to \Omega + 1 \qquad {\rm and} \qquad \Omega \to - 1/\Omega \ .
\end{equation}

In this respect, the bosonic model is a bit too simple. In proceeding to
the {\bf closed superstring}, one is forced to consider conformal fields
that are no more doubly periodic along the two homology cycles of the torus
(in modern terms, they are sections of line bundles, rather than
functions, on the torus). So, if one starts with proper ({\it i.e.}~doubly
anti-periodic, or Neveu-Schwarz) fermions, and if one insists on the
{\bf symmetries of a theory of closed strings} ({\it i.e.}~modular invariance),
one is forced to add more contributions. The way to do so is not unique, 
but a very suggestive possibility is to complete separately the contributions
of left and right movers ({\it i.e.}~the portions depending analytically and
anti-analytically on $\Omega$). The result is the GSO projection \cite{gso},
which leads to the ten-dimensional closed superstrings. So, at times one has 
to work harder to attain modular invariance, and this occurs precisely when
one is dealing with sections of nontrivial bundles.

It was the great contribution of \cite{orb} (see also \cite{orbs}) to
extend the consideration of sections of nontrivial bundles also to the
case of bosons. The resulting constructions, recognized as orbifolds of
toroidal models, have turned into a fundamental way of exploring the
structure of two-dimensional Conformal Field Theory. It is remarkable that
all known constructions based on free fields can be understood in these
terms. Indeed, this possibility for the closed superstrings was
pointed out in \cite{orb}, and served as a motivation for the orbifold
construction.

Armed with the principles of {\bf conformal invariance} and {\bf modular
invariance}, one can proceed to explore the set of conformal field theories.
Even restricting oneself to just twisting boundary conditions of free fields,
one finds a huge number of possibilities, and the long-standing limitations
on the dimensionality of space time  in String Theory fall apart \cite{4dim}.

The extent of the confidence on the properties of {\bf closed strings}, and the
corresponding amount of progress, have to be contrasted with the situation for
{\bf open strings}. Again, the content of the previous talks makes this point
hardly in need to be stressed. Apart from some occasional mention, open strings
have been completely left out. This is rather peculiar, because on the one 
hand String Theory was born in the form of the Veneziano amplitude for open 
strings, and on the other hand the resurgence of a wide interest in the 
subject was triggered by the discovery of the Green-Schwarz mechanism,
originally motivated by an analysis of open-string amplitudes. The excuses
that have been given over the times for the neglect of open strings can be
traced to two main points. First of all, the way symmetry groups originally
entered open string models is via the Chan-Paton ansatz \cite{cp}.
This consists in multiplying amplitudes by traces of suitable matrices (including those of the fundamental representations of the ${\rm O}(N)$, ${\rm U}(N)$ 
and ${\rm USp}(2N)$ Lie algebras \cite{class}). The whole thing looks rather ad
hoc, to be contrasted with the neat role played by Kac-Moody 
algebras \cite{km} in the construction of the heterotic string \cite{het}.
Moreover, the open (and closed) bosonic model is rather complicated, and thus
somewhat clumsy to deal with. It involves many more diagrams that the closed
(extended Shapiro-Virasoro) model, and often delicate divergence cancellations
between them. Furthermore, it is usually felt, not without regret, that modular
invariance is lost in this case, and that to check for the consistency of 
open-string models all one can do is appeal to anomaly cancellations, whenever
possible. This last point is made particularly dubious by the recent 
recognition that, in analogy with the special role played by the group 
${\rm SO}(32)$ in the type-I superstring in 10 dimensions, the group 
${\rm SO}(8192)$ selects a special bosonic model in 26 dimensions
\cite{dg,ms}.

The rest of this talk is devoted to remedying these inconveniences. My aim
will be convincing the reader, as well as I hope to have convinced the
listener, that {\bf the known open-string models in 26 and 10 dimensions
have to be understood as parameter-space orbifolds of corresponding 
left-right symmetric closed models}. The $Z_2$ twist involved mixes left
and right movers. In more geometrical terms, it symmetrizes between the
two sides of the parameter surfaces. So, the open Veneziano model in 26
dimensions is seen to descend from the closed (extended) Shapiro-Virasoro
model. The same pattern is followed in ten dimensions, where the type-I
superstring can be seen to descend from the chiral type-IIb superstring.
It should be noticed that the orbifold construction requires symmetry between
left and right waves. This means a chiral spectrum for the superstring, since
the two Ramond vacua must have the same chirality. The same point I will
try to make is that {\bf the size of the Chan-Paton group is determined
by modular invariance}, by which I mean that the orbifold projection,
applied to the surfaces with automorphisms that admit it, fixes the weights
of the diagrams, and explains the very emergence of open strings. {\bf
Open strings are the ``twisted'' sector of the construction}. Therefore,
their vertex operators sit at the fixed points. This is familiar stuff. After
all, we all knew for ages that open strings are emitted from boundaries!
The powers of two that build up the ``magic numbers'', 32 and 8192, can be
related to the sizes of the fixed-point sets. Actually, for these models,
Neil Marcus and I showed that the group theory can be generated by means of
free fermions valued on the boundaries of the parameter surfaces \cite{ms}.
This corresponds to the long-held picture of ``quarks at the ends of strings''.
From the point of view advocated here, one should keep in mind the analogy
with the zero modes of the Ramond sector in the orbifold construction of
the superstring.

Let me start by reviewing how Chan-Paton factors \cite{cp} originally
entered the game. The crucial observation was that
the cyclic symmetry of ``tree''
open-string amplitudes (what we would now call the ``disk contribution'')
is compatible with the multiplication by a trace of matrices. The cyclic
symmetry (planar duality) is the remnant, in the open-string case, of the
total symmetry (non-planar duality) of closed-string amplitudes. Demanding
that these (Chan-Paton) factors respect the factorization properties of 
amplitudes imposes severe constraints. These are somewhat relaxed if, following
Schwarz \cite{jhop}, one takes into account the twist symmetry, {\it i.e.}
the simple behavior of open-string amplitudes under world-sheet parity. The
result is an infinite number of constraints, solved long ago by Neil Marcus
and myself \cite{class}, by appealing to a classic result in the theory of
Algebras. This classifies the simple associative algebras over
the real numbers \cite{wed}. These arguments lead to exclude the exceptional
groups from the original open-string models in 26 and 10 
dimensions \cite{class}.

The Chan-Paton ansatz can actually be replaced by a dynamical construction
in terms of currents \cite{ms}, somewhat reminiscent of the corresponding
construction for the heterotic string. However, open string currents are
valued on the boundaries of the parameter surfaces. They are described, in
the simplest of ways, in terms of one-dimensional fermions, quantized 
anti periodically along boundaries. This implements the old intuitive idea
of ``quarks at the ends of strings'', but for a few subtleties. First of
all, in order to attain a degeneracy of order $2^{[D/2]}$, one needs but
$D$ ``quarks''. Second, these ``quarks'' have no space-time attributes.
Thus modifying the usual actions by adding 
\begin{equation}
S_{group} = \frac{i}{4} \, \int_{\partial \Sigma} \, ds \beta^I 
\frac{d \beta^I}{d s}
\end{equation}
produces all the right multiplicities for empty boundaries. A corresponding
modification of the vertex operators includes the $\beta$ fields, and produces
the trace factors. This is all good and well, but it would be nice to predict
the number of $\beta$ fields, especially since it turn out that one needs as
many of them as the string coordinates. Indeed, ten fermions give 
${\rm SO}(32)$, while 26 give ${\rm SO}(8192)$.

It has been known for a while that boundaries affect the conformal properties
of two-dimensional models. For instance, in ref. \cite{oalv} it is shown
that further divergences are introduced, proportional to the lengths of
boundaries, and that the theory ``feels'' the geodesic curvature of
boundaries. In ref. \cite{ms} we noticed that the ``smooth doubling'' of
surfaces forces the boundaries to be geodesics, which can be taken as a
boundary condition on the intrinsic metric. Thus, for each boundary, one is
left with a (non-logarithmic) divergence proportional to the length, with a
coefficient proportional to the number of space-time coordinates, and of the
right sign to cancel against the divergence introduced by the boundary
fermions. The divergence being not logarithmic, its cancellation is fraught
with ambiguities. Moreover, showing that the cancellation takes place requires
adding contributions coming partly from the parameter surface and partly from
its boundary. This involves standard ways of dealing with integrals of 
Dirac's delta functions over half of the real axis. This was all known to
Neil Marcus and myself at the time of writing \cite{ms}, but it is not
stressed there, since it can capture in different amounts one's interest,
due to the ambiguities mentioned above. Still, if one takes the cancellation
seriously, and if the boundaries are taken to be geodesics, 
{\bf the open-string models with proper groups
(${\rm SO}(32)$ and ${\rm SO}(8192)$) exhibit the same divergence structure
as closed string models.} Then, are they really to be regarded as
closed-string models all the way?
\vskip 15pt
\input epsf \centerline{ \epsfbox{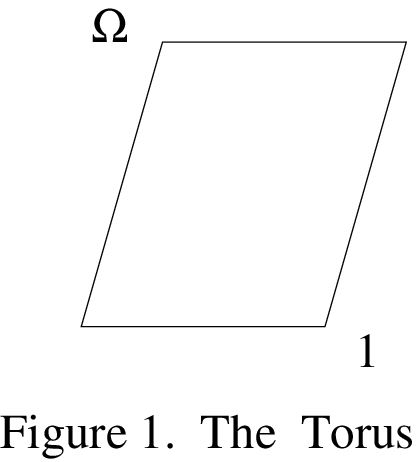}}
\vskip 10pt
A related observation is that the order of both ${\rm SO}(32)$ and
${\rm SO}(8192)$ is a power of two. More simply, empty boundaries contribute
a factor of two for each space-time coordinate. Such factors are familiar.
There are at least two instances where they arise. One is the Ramond sector
of the superstring. In this case one has gamma matrices and, after all, 
the quantization of the one-dimensional fermions of ref. \cite{ms} also
gives gamma matrices. The other case is apparently quite different. It is
the $Z_2$ orbifold of a ``square'' torus, described by Jeff Harvey at this
School. There the powers of two can be traced to the size of the fixed-point
set of the involution that defines the orbifold. This encourages one to look
for the same structure in the known open-string models in 26 and 10 dimensions.
\vskip 15pt
\input epsf \centerline{ \epsfbox{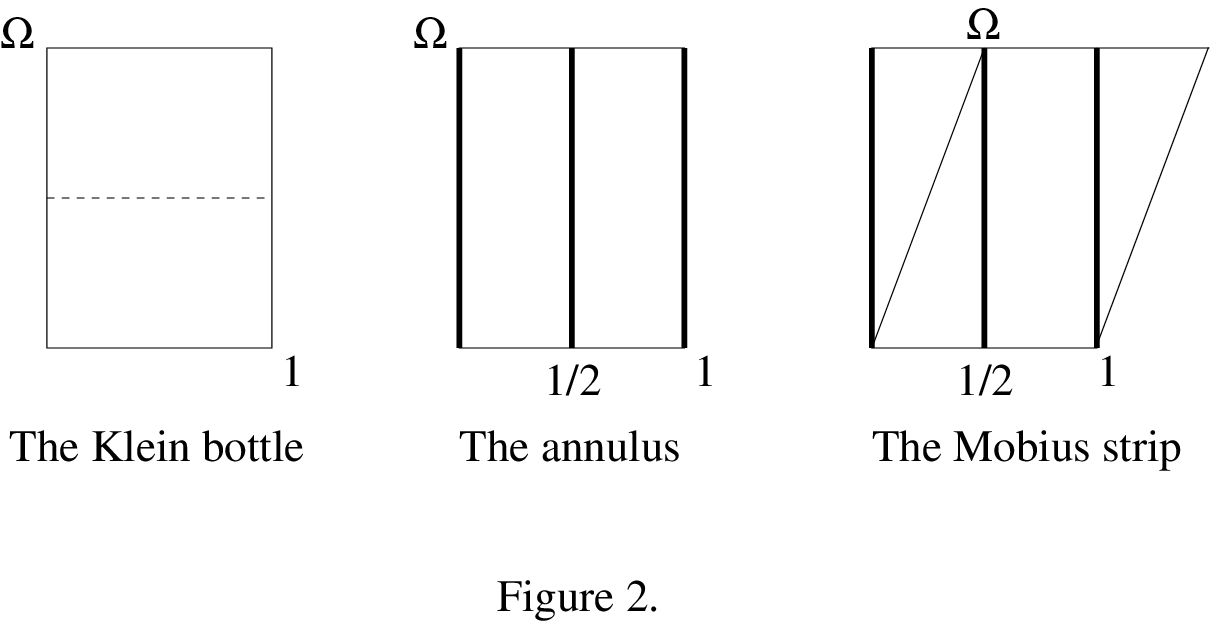}}
\vskip 10pt
As usual, it is simple and instructive to consider the genus-one contribution.
For simplicity, I will do so for the bosonic string in 26 dimensions. There
are then four diagrams, with parameter surfaces having, respectively, the
topology of the torus, the Klein bottle, the annulus and the M\"obius strip.
They can be conveniently described in terms of lattices in the complex plane
(figs. 1 and 2). In addition, the latter three surfaces are conveniently
described in terms of their ``doubles'', which are all tori \cite{ss}.
$\Omega$ is the ``period matrix'' of the doubles. It is purely imaginary
for both the Klein bottle and the annulus, but not for the M\"obius strip.
It should be noticed that the torus contribution is the same as for the closed
bosonic string, apart from a factor of two. Thus, it is modular invariant by
itself. Moreover, the Klein bottle is seen by inspection of figure 2 to be 
invariant under $\Omega \to \Omega + 2$. The resemblance between what one has
so achieved and the untwisted sector of the $Z_2$ orbifold described by Jeff
Harvey at this School is striking. The tricky point is that now the twist
affects the parameter surface. Thus, in looking for the analogue of the
$\Omega \to 1/\Omega$ transformation, one better think of what this 
transformation is meant to achieve. Then it is seen from figure 2 that the
annulus contribution is precisely what is needed, since the twist is rotated
by 90 degrees with respect to the Klein bottle. Finally, the M\"obius strip
symmetrizes the twisted sector. Actually, I have gone a bit too fast here. 
First of all, the two involutions that lead to the Klein bottle and to the
annulus are different. One results into two cross-caps, and the other into
two boundaries. Moreover, the ``shift'' in $\Omega$ that leads from the
annulus to the M\"obius strip is just $\frac{1}{2}$, not $1$. The clue is
noticing that the involutions act in the same way on the homology basis,
and this is all the string integrand ``feels''. On the other hand, the
zero modes result in the same contribution {\bf only if one works in terms of
the modulus of the double}. However, the ``proper time'' for the Klein
bottle is half the modulus of its covering torus, which is again half
the ``proper time''. Expressing amplitudes in terms of ``proper time''
exhibits the spectrum, and gives a relative factor of $2^D$ between Klein
bottle and annulus. This is the square of the multiplicity factor when the
gauge group has order $2^{[d/2]}$! In the same way, symmetrization in the
twisted (open) sector required adding $\frac{1}{2}$ to $\Omega$, which refers 
to the double, and generates precisely the modulus of the M\"obius strip.
The oscillator description accommodates the orbifold idea very nicely. Open
strings take values over ``one half'' of the parameter surfaces, and
{\bf are closed modulo the doubling of the parameter surfaces}.

Actually, the preceding discussion has left out an important point. This is
the choice of projection in the ground state of the twisted sector. In 
ref. \cite{ms} it was pointed out that a twist-even ground state, and thus 
the group ${\rm SO}(8192)$ rather than ${\rm USp}(8192)$, leads to the
elimination of some divergences via a principal part prescription. The
divergences manifest themselves in the small-$\Omega$ limit of the 
amplitudes corresponding to figure 2. We were inspired by a similar
phenomenon discovered by Green and Schwarz in the ${\rm SO}(32)$
superstring, and responsible for both finiteness and anomaly cancellation.
The last result in discussed at length in ref. \cite{prog}. Actually, even
for the superstring the same can be seen to occur directly at the level of
the partition function, if one refrains from using the ``aequatio'' of 
Jacobi, which sets to zero the contributions of the individual diagrams, due
to supersymmetry. The cancellations found by Weinberg \cite{weinb} in the
scattering amplitudes of the ${\rm SO}(8192)$ theory can be traced to the
same phenomenon.

The picture that emerges from the foregoing discussion has several facets.
One the one hand, it is particularly satisfying to see the structure of
Conformal Field Theory at work again. One the other hand, one looses the
need to consider open-string models as separate entities (or oddities). 
Everything fits into theories of closed strings, once one allows the 
possibility of {\bf twists mixing left and right-movers}, which have been
systematically avoided in discussions of orbifolds so far. The ``magic rule''
of modular invariance is recovered, and this should allow model building
with open strings as well. These points clearly deserve a fuller discussion,
which will be presented elsewhere.
\vskip 24pt
I am grateful to Giorgio Parisi and to the Organizers for making it possible
for me to come to Cargese. I am also grateful to Dan Friedan for a stimulating
discussion, and for his interest in the ideas presented in this talk.
Finally, the participants to the joint Mathematics-Physics seminar of the
two Universities of Rome, and in particular Massimo Bianchi, Emili Bifet
and Gianfranco Pradisi, contributed to making my stay in Rome both pleasant
and fruitful.

\end{document}